\documentclass[aip,jcp,amsmath,amssymb,floatfix,reprint,citeautoscript,articletitle=false,noeprint]{revtex4-1}

\usepackage{color}
\usepackage{graphicx} 
\usepackage{amsmath,amssymb,bm}
\usepackage[version=3]{mhchem}
\usepackage{float}
\usepackage{siunitx}
\usepackage[colorlinks,allcolors=black,citecolor=blue,urlcolor=blue]{hyperref}

\newcommand{\red}{{}}
\begin{document}
\title{%
Transferability of machine learning potentials:
Protonated water neural network potential applied to 
the protonated water hexamer
}
\author{Christoph Schran}
\email{christoph.schran@rub.de}
\affiliation{Lehrstuhl f\"ur Theoretische Chemie,
 Ruhr-Universit\"at Bochum, 44780 Bochum, Germany}
\affiliation{Department of Physics and Astronomy,
 University College London, London, WC1E 6BT, UK}
\affiliation{Present address: Department of Chemistry,
 University of Cambridge, Lensfield Road, Cambridge, CB2 1EW, UK}
\author{Fabien Brieuc}
\affiliation{Lehrstuhl f\"ur Theoretische Chemie,
 Ruhr-Universit\"at Bochum, 44780 Bochum, Germany}
\author{Dominik Marx}
\email{dominik.marx@rub.de}
\affiliation{Lehrstuhl f\"ur Theoretische Chemie,
 Ruhr-Universit\"at Bochum, 44780 Bochum, Germany}
\date{\today}

\begin{abstract}
A previously published neural network potential
for the description of protonated water clusters
up to the protonated water tetramer, \cf{H+(H2O)4},
at essentially converged coupled cluster accuracy 
(J.~Chem.~Theory~Comput. \textbf{16}, 88 (2020))
is applied to the
protonated water hexamer, \cf{H+(H2O)6}~--
a system that the neural network has never seen before. 
Although being in the extrapolation regime, it
is shown that the potential 
not only allows for
quantum simulations 
from ultra-low temperatures 
$\sim 1$~K 
up to
300\,K,
but that it is able to describe the new system very accurately
compared to 
explicit coupled cluster calculations. 
This
transferability of the model is rationalized
by the similarity of the atomic environments encountered
for the larger 
cluster 
compared to the environments
in the training set of the model.
Compared to the interpolation regime
the quality of the model is reduced by roughly one
order of magnitude, but most of the
difference to the coupled cluster reference
comes from global shifts of the potential energy surface,
while local energy fluctuations are well recovered.
These results suggest that the application of 
neural network potentials in extrapolation regimes
can provide 
useful results and 
might be more
general than usually thought.
\end{abstract}
\maketitle

\section{Introduction}
\label{sec:intro}

In recent years, machine learning has become
a compelling tool for the representation of
potential energy surfaces~\cite{Behler2016/10.1063/1.4966192,
Bartok2017/10.1126/sciadv.1701816,
Butler2018/10.1038/s41586-018-0337-2,
Deringer2019/10.1002/adma.201902765,
Mueller2020/10.1063/1.5126336,
Manzhos2020/10.1021/acs.chemrev.0c00665}.
The first family of such machine learning potentials based on
artificial neural networks 
which scale
to arbitrary system sizes
were high-dimensional neural network
potentials~\cite{Behler2007/10.1103/PhysRevLett.98.146401,
Behler2017/10.1002/anie.201703114}
(NNPs).
Since then, many distinctly different approaches either also based
on artificial neural networks~\cite{%
Ghasemi2015/10.1103/PhysRevB.92.045131,%
Khorshidi2016/10.1016/j.cpc.2016.05.010,%
Artrith2017/10.1103/PhysRevB.96.014112,%
Smith2017/10.1039/C6SC05720A,%
Gastegger2017/10.1039/c7sc02267k,%
Unke2019/10.1021/acs.jctc.9b00181,%
Shao2020/10.1021/acs.jcim.9b00994%
}
or on 
kernel methods~\cite{
Bartok2010/10.1103/PhysRevLett.104.136403,%
Rupp2012/10.1103/PhysRevLett.108.058301,%
Thompson2015/10.1016/j.jcp.2014.12.018,%
Shapeev2015/10.1137/15M1054183,%
Li2015/10.1103/PhysRevLett.114.096405,%
Chmiela2017/10.1126/sciadv.1603015%
}
have been introduced over the years,
while recent development following the principles
of deep-learning has allowed one to incorporate
parts of
the description
of the chemical environments in the architecture
of the model~\cite{
Schuett2017/10.1038/ncomms13890,%
Zhang2018/10.1103/PhysRevLett.120.143001%
}.
While there is usually agreement in the community 
that these models can only be used in order to interpolate
between a meaningful set of training points,
there have been recent examples that show a broader
transferability
of such machine learning approaches
than previously assumed.
This includes, for example, the application of
NNPs
\cite{Behler2007/10.1103/PhysRevLett.98.146401,
Behler2017/10.1002/anie.201703114}
for
alkanes to larger chains~\cite{Gastegger2016/10.1063/1.4950815}
and
for
liquid water to various ice phases~\cite{Monserrat2020/10.1038/s41467-020-19606-y}
as well as a
Gaussian approximation potential~\cite{Bartok2010/10.1103/PhysRevLett.104.136403}
for carbon to random structure searches~\cite{Rowe2020/10.1063/5.0005084}
that explored quite different configurations than the diverse carbon phases
used to train the model.
In this communication we show that
such generalization capability
can also be obtained for protonated water clusters
of 
larger size than the clusters used in 
the training set
of the machine learning model.
For that purpose we use our previously 
published NNP
(exactly as reported in the Supporting Information in 
Ref.~\citenum{Schran2020/10.1021/acs.jctc.9b00805}) 
for the description of protonated water clusters,
which has been developed in an automated and adaptive
workflow.
It has been trained on essentially converged coupled cluster reference
data for 
protonated water 
clusters 
up to 
\cf{H+(H2O)4} 
--- including also the water monomer itself.
This 
implies that the 
very same
model is able to describe the potential
energy surface of all differently sized clusters 
\cf{H+(H2O)$_n$}
(here 
from $n=1$ up to~$4$) 
on equal footing 
by virtue of including all of them explicitly and simultaneously in the training. 
In contrast, 
conventional
many-body expansions,
which have also been successfully applied to protonated water
clusters~\cite{Pinski2014/10.1021/ct400488x},
expand the potential energy as a sum of many-body \textit{corrections}
to the previous terms.
Here, we apply our model to 
path integral molecular dynamics (PIMD) 
quantum simulations
of the protonated water hexamer, \cf{H+(H2O)6},
in its extended Zundel conformation
(i.e. the hydrated protonated water dimer), 
a system that 
was not included in training the NNP.
The structures generated by these stable quantum
simulations are afterwards validated with respect
to single-point coupled cluster calculations of the
same quality as those used to generate the NNP up to 
$n=4$
only. 
We show that 
this particular
model is able to provide
meaningful and accurate predictions in
this extrapolation regime.
These promising results are rationalized
by the similarity of the atomic environments
encountered
in this extrapolation regime
to the 
ones obtained for 
the smaller protonated water clusters
present in the training set of the model.

\begin{figure}
   \centering
   \includegraphics[scale=1]{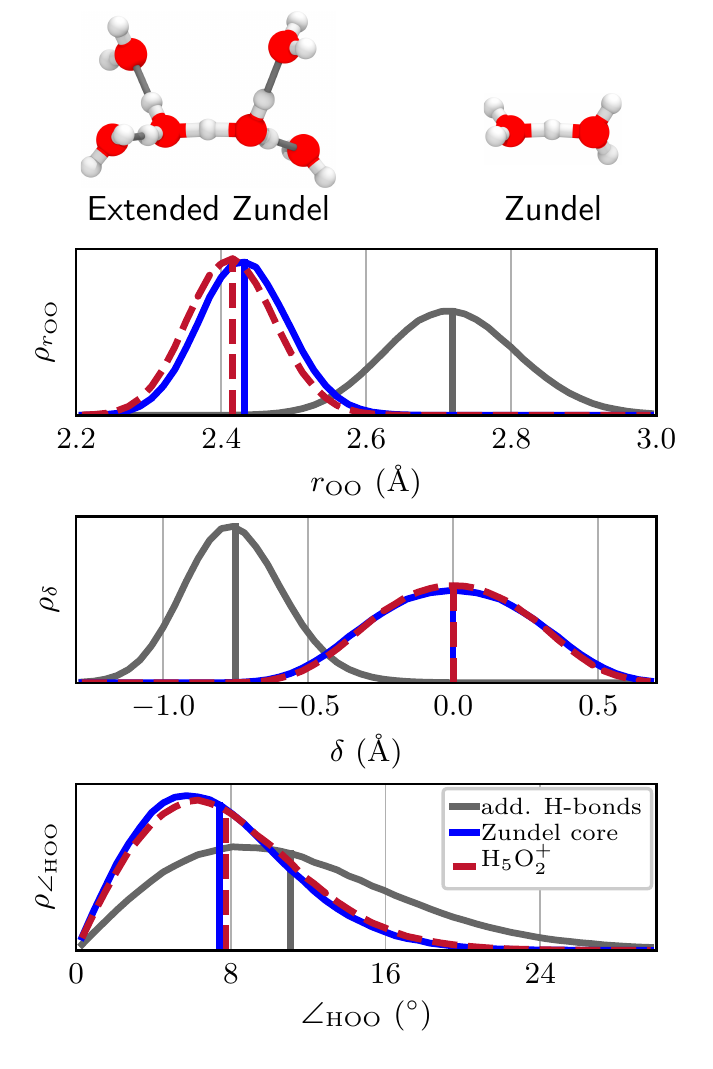}
   \caption{
Normalized probability distributions of the heavy atom
donor-acceptor distance $r_\text{OO}$ (top),
the proton-sharing coordinate $\delta$ (middle) defined
as $\delta=r_\text{OH}-r_{\text{H}\cdots\text{O}}$,
and the hydrogen bond angle $\angle_\text{HOO}$ (bottom)
from PIMD simulations at 1.67\,K for the central,
ultra-strong hydrogen bond (blue) and the 
additional
four
hydrogen bonds to the dang\-ling water molecules (gray)
in the extended Zundel cation, \cf{(H5O2+)(H2O)4},
as well as for the smaller 
bare 
Zundel cluster, \cf{H5O2+}, (red)
for comparison.
The average of the distributions is marked 
with vertical lines using the same color code. 
Only
hydrogen bonded configurations are considered
in this analysis based on 
a standard hydrogen bond criterion~\cite{Luzar1996/10.1103/PhysRevLett.76.928}.
The optimized minimum energy structures of the two considered
systems are displayed on top of the figure.
}
   \label{fig:struc}
\end{figure}

\section{Results and Discussion}
\label{sec:results}
In order to test the application of the previously
published NNP for the description of protonated water clusters
beyond what was considered in the original
development~\cite{Schran2020/10.1021/acs.jctc.9b00805},
we ran path integral simulations at various temperatures
starting from the optimized minimum energy structure
of the extended Zundel cation.
The computational details of these simulation are described
in the appendix.
All tested quantum simulations from 
close to the ground state
at 1.67\,K up to 100\,K 
were stable~--
although all being evidently in the extrapolation regime
of the NNP. 
The model was indeed 
exclusively
extrapolating since for
each and every 
configuration
encountered during the simulations the associated values of
the atom-centered symmetry functions~\cite{Behler2011/10.1063/1.3553717},
used here as descriptors of the atomic environments, 
were outside the range of values
present
during training.
In particular, we mainly observe extrapolation of 
the two broadest radial
symmetry functions,
one centered 
around oxygen atoms and 
the other
one
centered around hydrogen atoms,
which both involve 
distant oxygen atoms
in their pairs.
This
indicates that
exclusively
unknown configurations
were encountered
in all steps of these simulations. 
While simulations up to 100\,K sampled only the
extended Zundel isomer, \cf{H5O2+(H2O)4},
we applied the model also at
temperatures of 200, 250, and 300\,K, which
resulted in occasional rearrangements of the cluster to
other known minima.
Out of the 12 simulations at higher temperatures
two runs rearranged into strained four-membered ring structures
for which the NNP provided unphysical predictions
after about 200\,ps simulation time,
a phenomenon we analyze in more detail towards the end
of this communication.
This mostly stable application of the NNP in 
a
extrapolation regime is a first promising indication that the potential
could be applied beyond the originally considered
cluster sizes.
In Fig.~\ref{fig:struc} we show the probability
distribution functions of three main structural
properties of the hydrogen bond~\cite{Schran2019/10.1039/C9CP04795F}
from the simulation
of \cf{(H5O2+)(H2O)4}
at 1.67\,K and compare them to the respective distributions
of the much smaller bare Zundel cation, \cf{H5O2+}.
As seen therein, the model provides the expected
bimodal
distributions of the donor-acceptor distance $r_\text{OO}$
and the proton-sharing coordinate $\delta$
for the extended Zundel cation, caused by the two distinctly
different types of hydrogen bonds in the system:
The central ultra-strong hydrogen bond (shown in blue)
and the four 
additional
hydrogen bonds to the dangling water molecules (shown in gray).
In comparison, the bare Zundel complex has 
qualitatively similar 
distributions 
in case of
the central hydrogen bond, but 
features a distinct shift towards 
shorter 
donor-acceptor distances
as well as a slightly more localized proton-sharing coordinate. 
Finally, the distribution of the HOO~angle for the central, ultra-strong
hydrogen bond in the extended Zundel cation is again
close to the one in the bare Zundel complex, 
while the four weaker hydrogen bonds to the dangling water
molecules feature the expected broader distribution
that is shifted to larger angles.
These results
are 
in substantial
agreement
with previous studies on the extended Zundel cation~\cite{
Jiang2000/10.1021/ja990033i,
Christie2002/10.1021/jp0209042,
Nguyen2009/10.1021/ct900123d,
Heine2013/10.1021/ja401359t,
MouhatPhDthesis,
Heindel2018/10.1021/acs.jctc.8b00598}
regarding the symmetric nature of the central ultra-strong
hydrogen bond and 
its
close match with 
that in 
the smaller bare Zundel cation.
This
highlights the physically meaningful nature
of the quantum structures generated with our model
operating here in extrapolation mode,
even
close to the quantum ground state  
at 1.67\,K.
\begin{figure*}
   \centering
   \includegraphics[scale=1]{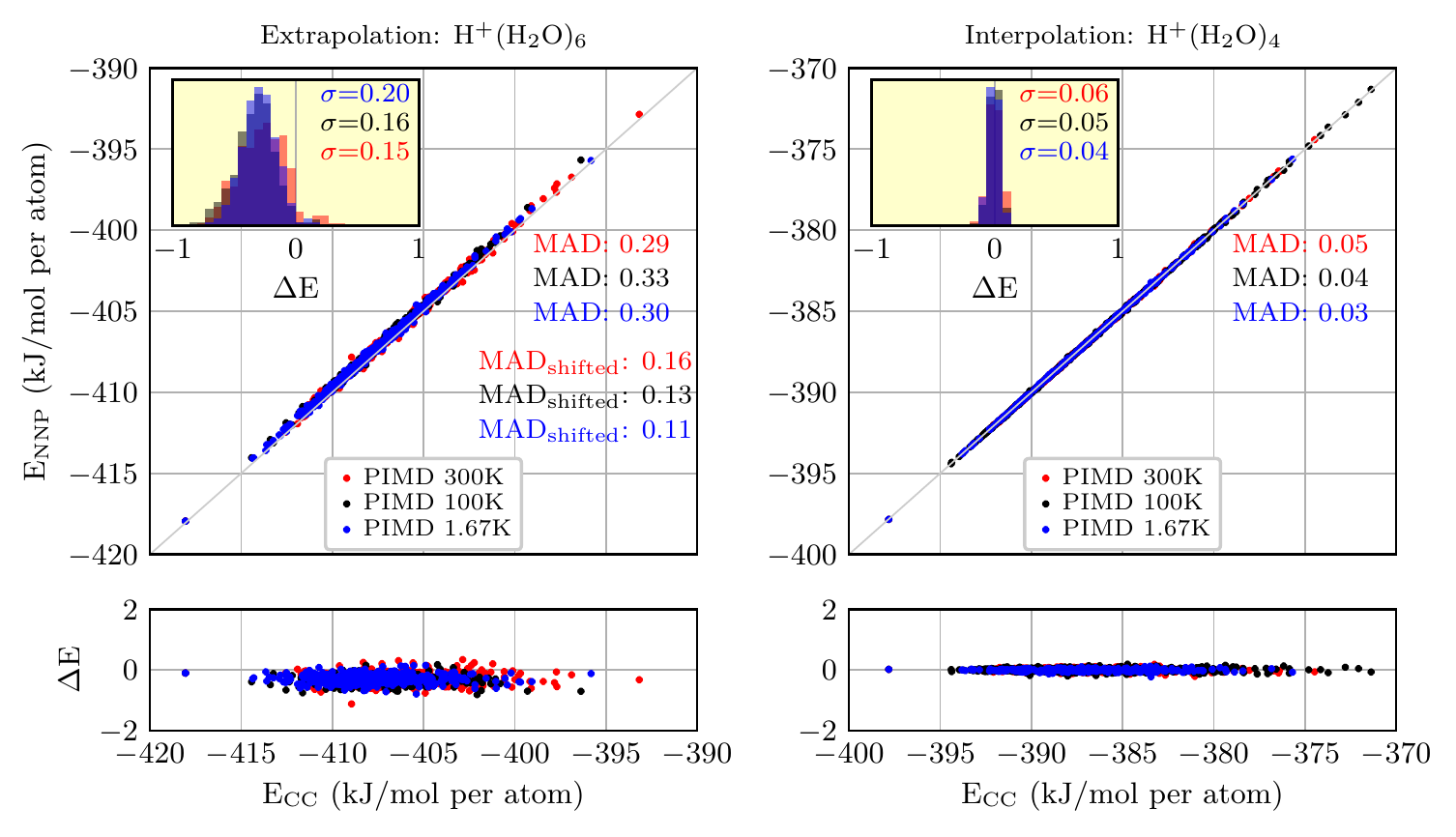}
   \caption{
Correlation of the energy per atom from explicit
CCSD(T*)-F12a/aug-cc-pVTZ calculations (CC) and the
NNP predictions for 300 randomly selected
configurations of the protonated water hexamer
(left panels)
and of the protonated water tetramer
in its Eigen conformation (right panels)
at 1.67 (blue), 100 (black),
and 300\,K (red),
respectively.
The mean absolute difference (MAD) for all temperatures
are reported in
their respective color.
In the upper left pannel we also include the shifted MAD
values, 
see text, for which the 
systematic 
bias 
(as quantified in the upper left inset)
is removed from the NNP prediction by
shifting the energies by the
MAD at the respective temperature
(but without showing the underlying shifted data themselves). 
The lower panels show  the  energy  differences between
CC
reference  and  NNP  prediction  over
the  whole  range  of  reference energies,  while  the
inset  in  the  upper  panels  show  the  histograms
of the energy differences including the corresponding
standard deviations $\sigma$ in the respective color.
}
   \label{fig:nnp-random-validation}
\end{figure*}
In a next step, the quality of the NNP prediction 
during the simulations is validated 
by explicitly
evaluating the coupled cluster reference method
(namely CCSD(T*)-F12a/aug-cc-pVTZ, 
see the appendix for details)
for 300~randomly selected configurations at
1.67, 100, and 300\,K.
We note that due to the system size
such coupled cluster 
calculations are increasingly demanding
in view of their steep scaling 
not only in terms of computation time, but also 
when it comes to
memory resources.
This growth is 
such that it would have been challenging to explicitly include the
protonated water hexamer when training the NNP in the first place.
The resulting correlation of the NNP prediction
and the actual coupled cluster reference 
for \cf{H+(H2O)6}
is shown in the left panels of Fig.~\ref{fig:nnp-random-validation}.
In the same figure we include in the right panels the equivalent
test for
the largest cluster considered during the
development of that NNP, namely the protonated water tetramer,
\cf{H+(H2O)4},
in its Eigen structure 
(corresponding to the hydrated hydronium cation~/ protonated water monomer
\cf{H3O+}).
Note that
the same energy scales
are used
for all axes to allow for one-to-one comparisons. 
Overall, a high correlation between reference and
NNP is obtained, which is especially remarkable
for a NNP that operates exclusively in the extrapolation
regime as 
is the
case for the
protonated water hexamer
(left panels).
The 
application of NNPs
in interpolation regimes,
as demonstrated here for \cf{H+(H2O)4} (right panels), 
is evidently yielding
better results, but the quality of
the NNP for the
larger 
and thus fully extrapolating
system is only reduced
by about one order of magnitude and~-- most
importantly~-- does not feature 
strong outliers
for all 300~configurations considered in this cross-check. 
In addition, it can be seen that the prediction
of the NNP in the extrapolation regime is affected
by a systematic bias as evident from the shift of the
histogram of the energy differences 
w.r.t. zero 
shown in the inset of the upper left panel.
If this bias is removed from the prediction of the NNP by 
uniformly 
shifting 
all
energies by the respective mean absolute
difference (here 0.30, 0.33
and 0.29~kJ/mol per atom at 300, 100 and 1.67\,K, respectively),
the precision of the prediction is 
improved roughly by a factor of three
at no additional cost. 
We note in passing that we observe essentially the same
quality for the low temperature simulations,
that sample exclusively the extended Zundel conformer,
\red{
as
}
for the configurations at 300\,K,
although rearrangements to other known minima occur
under these conditions.

\begin{figure*}
   \centering
   \includegraphics[scale=1]{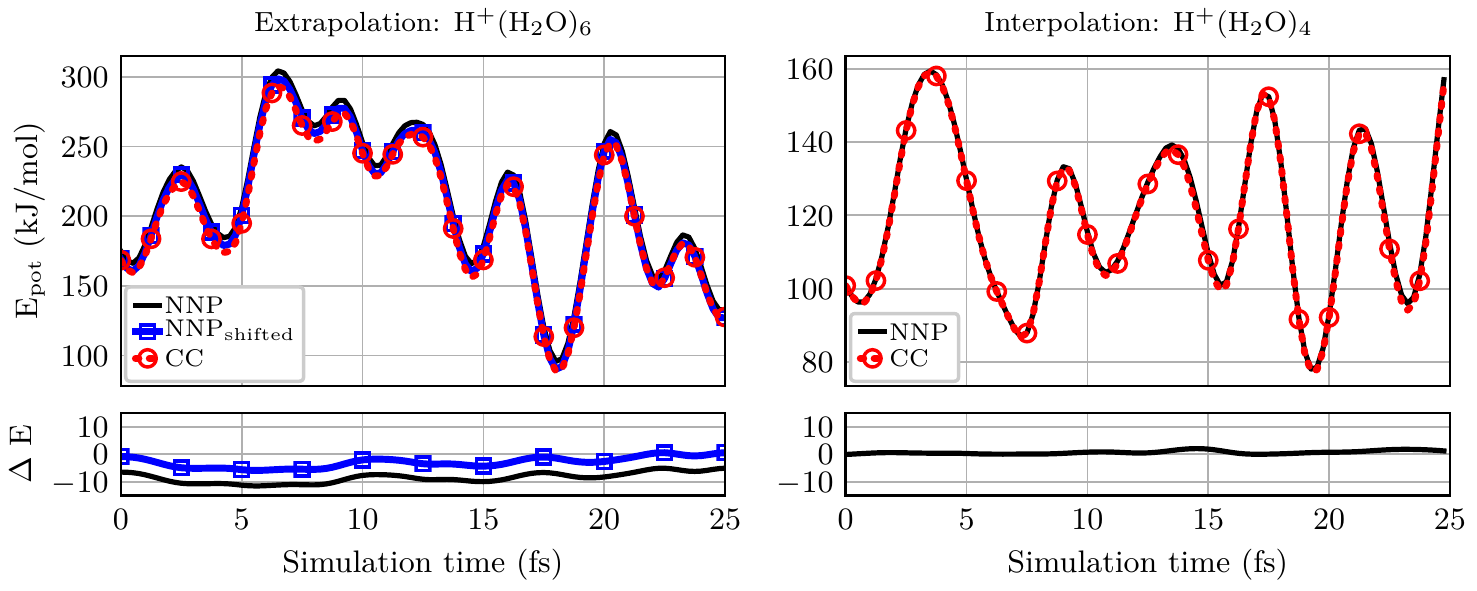}
   \caption{
Potential energy along one replica of quantum PIMD trajectories at 1.67\,K 
of the extended Zundel cation (left) and the protonated
water tetramer in the Eigen conformation (right) using the 
original and shifted (see text) neural network potentials 
(NNP, NNP$_{\rm shifted}$ only in the left panel).
The 
CCSD(T*)-F12a/aug-cc-pVTZ reference (CC)
was obtained by recomputing
the energies 
for each configuration
along the NNP trajectories and are shown as red dotted
lines (with only a few circles added since the CC energies mostly superimpose the NNP data).
These 
energies are reported relative to the respective equilibrium structure.
The bottom panels 
highlight
the respective energy differences
of the NNP predictions 
to the CC reference method.
   }
   \label{fig:nnp-md-validation}
\end{figure*}
To further validate the application
of the NNP for the extended Zundel cation
when used in realistic simulations, 
we additionally re-evaluated a short segment of
the trajectory associated with one replica,
or bead, of the quantum path integral 
molecular dynamics simulations 
at 1.67\,K with the coupled cluster
reference 
method.
The resulting potential energy profile along
these 25~fs of the simulation 
is shown in the left panel of Fig.~\ref{fig:nnp-md-validation}.
We also carried out the same analysis
for the largest cluster
explicitly considered in the construction of the NNP, 
the protonated water tetramer, 
in the right panel to allow for one-to-one comparison 
to the interpolation regime. 
As before, the NNP yields surprisingly good
results for the extended Zundel cation, although
working exclusively in its extrapolation mode. 
The NNP is 
found to be
able to recover the overall energy
profile along the short segment of the trajectory
and correctly reproduces the energy fluctuations
(black line). 
As already seen for the randomly selected
structures, the NNP is affected by 
an overall
bias that shifts the prediction to slightly larger energies.
However, if we use the mean absolute difference
of the 300~random structures from the simulation
at 1.67\,K to shift the energy prediction according to the NNP, 
the coupled cluster reference energy profile is recovered 
perfectly by the NNP$_{\rm shifted}$ on the physically relevant scale as set by the potential energy fluctuations
(blue line with squares). 
As known from the originally published benchmarking
of the NNP~\cite{Schran2020/10.1021/acs.jctc.9b00805},
the Eigen cation does not suffer from such a bias
since it has been used explicitly to train that NNP 
as shown in the right panel of Fig.~\ref{fig:nnp-md-validation}.
Overall, this analysis reveals that the 
fully extrapolating NNP 
is able to 
almost perfectly
recover the correct energy fluctuations
of the extended Zundel cation
even in the deep 
quantum
regime at 1.67\,K close to the ground state
after correcting for the global energy shift.
Even without any such shift of the energies,
the NNP performs unexpectedly well in such an
extrapolation regime, 
which suggests that there can be great potential in 
exploiting
extrapolation capabilities for building
more complex machine learning potentials.
Let us finally provide some insight into the unexpected
transferability of the model to the protonated water hexamer.
In order to check whether the predictive power of
our model is the sole result of the similarity to
the protonated water tetramer, we have trained
a separate NNP only to the tetramer configurations
present in the training set.
This model could not be successfully applied in
PIMD simulations as rearrangements into
unphysical configurations are observed at all
temperatures after some short initial period.
We thus conclude that the smaller clusters grant
additional robustness to the model and contribute to
its transferability.

\begin{figure}
   \centering
   \includegraphics[scale=1]{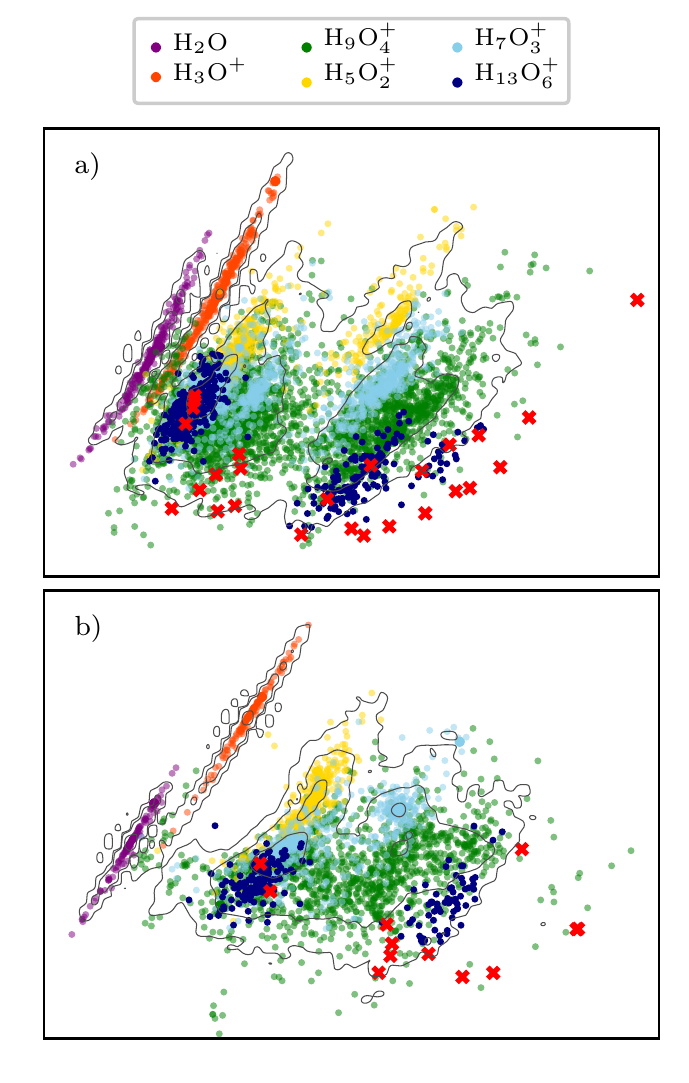}
   \caption{
Two dimensional projection based on the principle
component analysis of the hydrogen (a) and
oxygen (b) atomic environments as encoded by the
atom centered symmetry functions of the training set
of the NNP model.
\red{
The differently sized clusters
for a subset of the atomic environments in the training set
are marked by different colors, while the contour
lines show the
probability density of the PCA projection
for all atomic environemts in the training set
in equidistant steps on a logarithmic scale.
}
The same projection, shown in dark-blue, is also applied to the
configurations of the protonated water hexamer
as generated by PIMD simulations at various temperatures.
Red crosses mark the atomic environments of
two configurations of the protonated water hexamer
which feature unphysical predictions of the NNP model.
\label{fig:pca}
}
\end{figure}

To further understand this behavior,
we have analyzed the similarity of
the atomic environments encountered in the extrapolation
regime, as encoded by the atom centered symmetry functions,
to the environments in the training set of the model.
For that purpose, we performed a principle component analysis, separately
for the oxygen and hydrogen environments, and projected
the descriptors onto the two most relevant components.
These two components together 
account for 72\% and 82\% of the variance
in the full dimensional feature space of oxygen
and hydrogen environments,
respectively.
Subsequently, the same projection is applied to the
configurations of the protonated water hexamer
as depicted in Fig.~\ref{fig:pca}.
As 
can be
seen 
from the dark-blue points in this figure,
the configurations of the protonated water
hexamer remain within the boundaries of the training set
for the most relevant components of the descriptor space,
\red{
although we have observed
}
extrapolation for a subset of the descriptors.
This observation is in line with other recent studies
that have reported on the generalization capabilities
in machine learning models for various systems~\cite{Gastegger2016/10.1063/1.4950815,
Rowe2020/10.1063/5.0005084,
Monserrat2020/10.1038/s41467-020-19606-y}.

In addition, we have also included the projection of
the atomic environments in the two higher temperature
conformers that feature strained four-membered rings
and were observed to lead to unphysical predictions
of the model, 
see the red crosses in Fig.~\ref{fig:pca}.
Clearly, the main components of these atomic environments
leave the region spanned by the training set, which thus
can be associated with the missing predictive power
of the model.
This points towards the limitations of the
application of machine learning models in extrapolation
regimes, as sufficient similarity to the training set
is required to provide meaningful results.
We note that these limitations can in principle be overcome
by continuing the automated fitting process of the model
and explicitly targeting the larger cluster by selecting
the most representative configurations from our
exhaustive simulations.

\section{Conclusion and Outlook}
\label{sec:conclusion}
In summary, our 
previously published neural network potential,
trained for simulations up to the protonated water tetramer,
performs unexpectedly well for the extended Zundel cation,
a conformer of the protonated water hexamer.
This is
notable
since these simulations
are carried out entirely in the extrapolation regime,
which means that only
unknown configurations
are encountered
in each and every simulation step. 
This NNP
does not only allow one to run stable 
path integral
quantum simulations,
but also recovers correctly
the 
quantum-thermal
energy fluctuations
in direct comparison to the coupled cluster reference data
of the extended Zundel complex. 
Yet, the NNP suffers from a slight global shift of the
potential energy, which however does not influence the 
relative fluctuations. 
Moreover, we show how this shift can be systematically corrected
\textit{a~posteriori} based on rather few additional reference calculations
of the extended complex.

These promising results are explained 
by the similarity of the atomic environments
encountered
in the present application 
with the environments
of the smaller clusters that composed
the training set 
of the model,
which accurately treats a class of molecular complexes consisting of three up to 
13~constituting atoms. 
Since the training set of the neural network potential is composed of
clusters from 
the hydronium cation up to the protonated water tetramer
and does include the water monomer as well, 
sufficient local environments are 
considered in its construction 
to yield these reassuring results.

Still, 
we would like to stress that these promising
results should not be taken for granted and 
using 
machine learning models beyond their considered
scope of application 
during their construction and parameterization 
requires 
great care and
sufficient 
validation.
Evidently, in most cases that are too far away from the chemical
space spanned by the training set, machine learning
models will 
provide unphysical results.
We 
observed
this in the present case
at higher temperatures, where significant topological
rearrangements
can
occur that generate distinctly
different atomic environments in the
protonated water clusters
compared to those covered by
our training set.
At the same time, our results reveal that there
can be cases
where
carefully constructed machine learning
models are applicable 
much 
beyond interpolation regimes.
This could be a promising route for the development
of more complex machine learning models.
Thus, we hope that our short report will stimulate methodological work 
to explore systematically and fundamentally the power of
machine learning models to safely conquer unknown territory.

\begin{acknowledgments}
We are thankful to Harald Forbert,
Ondrej Marsalek, and Venkat Kapil
for insightful discussions.
Funded by the \textit{Deutsche Forschungsgemeinschaft}
(DFG, German Research Foundation) under Germany's
Excellence Strategy~-- EXC 2033~-- 390677874~-- \mbox{RESOLV}.
C.S. acknowledges partial financial support from the 
\textit{Alexander von Humboldt-Stiftung}.
The computational resources were provided by 
HPC@ZEMOS, HPC-RESOLV, and BoViLab@RUB.
\end{acknowledgments}

\section*{Data availability}
The data that supports the findings of this study
are available within the article.

\appendix
\section*{Appendix: Computational details}
\label{sec:comp-det}

Path integral molecular dynamics simulations of the extended Zundel cation
down to \SI{1.67}{\kelvin} have been performed with the 
\texttt{CP2k} program package~\cite{CP2K,Hutter2014/10.1002/wcms.1159}.
The potential energy surface was described
using a recently developed and published~NNP
fitted to coupled cluster reference calculations~\cite{Schran2020/10.1021/acs.jctc.9b00805}.
This NNP describes all 
protonated water clusters, from the 
protonated water monomer (hydronium cation) up 
to the protonated water tetramer considered in the development, 
on equal footing
and, in particular, also explicitly includes the water monomer.
It has been shown to not only match the reference coupled cluster
theory with very high precision, but is also able to accurately
describe proton transfer in the considered clusters~\cite{Schran2019/10.1039/C9CP04795F}.
We apply this NNP~\cite{Schran2020/10.1021/acs.jctc.9b00805}
in the present study to the larger
protonated water hexamer in its
extended Zundel 
structures, 
which is thus entirely in the
extrapolation regime of the model.

The extended Zundel cation was simulated at temperatures of
300, 250, 200, 100, 20, 10 and \SI{1.67}{\kelvin}
including the quantum nature of the nuclei
using the PIQTB
thermostat~\cite{Brieuc2016/10.1021/acs.jctc.5b01146}
which has been recently extended to 
and validated at
ultra--low temperatures~\cite{Schran2018/10.1021/acs.jctc.8b00705}.
In order to reach convergence, the path integral was
discretized using $P=$~6, 8, 12, 16, 64, 128, and 256
replica at $T=$~300, 250, 200, 100, 20, 10, and \SI{1.67}{\kelvin},
respectively. 
The convergence of these path integral discretizations 
was chosen according to 
previous
explicit benchmarking results
\cite{Schran2018/10.1021/acs.jctc.8b00705}
for the prototypical hydrogen bond in the bare Zundel cation.
For comparison, we also performed simulations using exactly the same settings
for the Eigen and bare Zundel clusters,
which were explicitly considered in the construction of the model~\cite{Schran2020/10.1021/acs.jctc.9b00805}.
All reported simulations were propagated
in four independent runs
for 
in total
\SI{1}{\nano\second}
using a formal molecular dynamics
time step of \SI{0.25}{\femto\second},
while \SI{10}{\pico\second} at the beginning of each
simulation were discarded as equilibration.

Explict validations of the 
predictive power 
of the NNP
which is used here exclusively in its extrapolation regime 
was achieved by reevaluating the energies 
of many configurations of the extended Zundel complex
with the same coupled cluster method
as used for the development of that NNP.
These calculations of the CCSD(T) reference energies 
were performed with the
\texttt{Molpro} program package\cite{MOLPRO}
by employing the explicitly correlated
F12a method~\cite{Adler2007/10.1063/1.2817618,
Knizia2009/10.1063/1.3054300}
to correct for the basis set incompleteness error.
As suggested~\cite{Knizia2009/10.1063/1.3054300} 
we additionally employed the
size-consistent scaling of the perturbative triples, (T*), 
together with the aug-cc-pVTZ basis
set~\cite{Dunning1992/10.1063/1.462569,
Woon1994/10.1063/1.466439}.
This so-called
CCSD(T*)-F12a/aug-cc-pVTZ
electronic structure setup 
has been shown to provide energies
very 
close to the complete basis set (CBS) limit~\cite{Knizia2009/10.1063/1.3054300}.
\end{document}